# AUDIO TAGGING WITH CONNECTIONIST TEMPORAL CLASSIFICATION MODEL USING SEQUENTIAL LABELLED DATA

**Yuanbo Hou[1], Qiuqiang Kong[2] and Shengchen Li[1]**

**Abstract.** Audio tagging aims to predict one or several labels in an audio clip. Many previous works use weakly labelled data (WLD) for audio tagging, where only presence or absence of sound events is known, but the order of sound events is unknown. To use the order information of sound events, we propose sequential labelled data (SLD), where both the presence or absence and the order information of sound events are known. To utilize SLD in audio tagging, we propose a Convolutional Recurrent Neural Network followed by a Connectionist Temporal Classification (CRNN-CTC) objective function to map from an audio clip spectrogram to SLD. Experiments show that CRNN-CTC obtains an Area Under Curve ($AUC$) score of 0.986 in audio tagging, outperforming the baseline CRNN of 0.908 and 0.815 with Max Pooling and Average Pooling, respectively. In addition, we show CRNN-CTC has the ability to predict the order of sound events in an audio clip.

## 1.1 Introduction

Audio tagging aims to predict an audio clip with one or several tags. Audio clips are typically short segments such as 10 seconds of a long recording. Audio tagging has many applications in information retrieval [1], audio classification [2], acoustic scene recognition [3] and industry sound recognition [4].

 Many previous works of audio tagging relies on strongly or weakly labelled data. In strongly labelled data [3], each audio clip is labelled with both tags and onset and offset times of sound events. Labelling the strongly labelled data is time consuming

---

[1] Y. Hou (✉) · S. Li
Beijing University of Posts and Telecommunications, Beijing, China
e-mail: hyb@bupt.edu.cn

[2] Q. Kong
Centre for Vision, Speech and Signal Processing, University of Surrey, UK

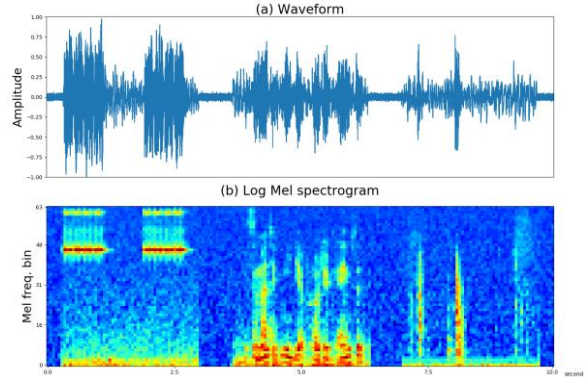

strong labels:
alert — noise — speech — noise — pageturn

sequential labels: (alert, speech, pageturn)

weak labels: (speech, alert, pageturn) or (speech, pageturn, alert) or (pageturn, alert, speech)

**Fig. 1.1** From top to bottom: (a) waveform of an audio clip containing three sound events: "alert", "speech" and "pageturn"; (b) log Mel spectrogram of (a); Strong labels, sequential labels and weak labels of the audio clip.

and labor expensive, so the size of strongly labelled dataset is often limited to minutes or a few hours [5]. Additionally, the onset and offset time of some sound events are ambiguous due to the fade in and fade out effect [6]. On the other hand, many audio datasets contain only the tags, without the onset and offset times of sound events. This is referred to as weakly labelled data (WLD) [7]. Many audio tagging dataset are weakly labelled [2, 6] and are often larger than strongly labelled datasets [3, 5]. However, in WLD, only the presence or absence of sound events are known, the occurrence sequence of sound events are not known. These weakness limit the use of strongly labelled data and weakly labelled data.

To avoid the weakness of strongly labelled data and WLD and use order information of sound events, we propose sequential labelled data (SLD). This idea is inspired by the label sequences in speech recognition [8]. In SLD, the tags and order of tags are known, without knowing occurrence time of tags. SLD not only reduces the workload of data annotation and avoids the problem of inaccurate time positioning of tags in strongly labelled data, but also indicates the order of tags in WLD. Compared with strong tags, there is no occurrence times of tags in SLD. Compared with weak tags, the order of tags is known in SLD. In addition, the order information of events will benefit tasks like acoustic scene analysis [3] and environment recognition [4]. Fig. 1.1 shows an audio clip and its strong, sequential and weak tags.

To utilize the SLD in audio tagging, we propose to use CTC technique to train a CRNN (CRNN-CTC). CTC is a learning technique for sequence labelling with RNN [9], which has achieved great success in speech recognition [8]. In fact, CTC is an objective function that allows RNN to be trained for sequence-to-sequence tasks, without requiring any prior alignment between the input and target sequences [8]. In training, CTC computes the total probability of input sequences, sums over all possible alignments [9]. CTC allows train an RNN without any prior alignment (i.e. the starting or ending times of each sound event), hence, even without strong labels, it is sufficient to do audio tagging with SLD based on CTC model, the details will be described in section 1.4.

There are two contributions in this paper. First, in audio tagging, we propose SLD, which not only reduces the workload and difficulties of data annotation in strong labels, but also indicates the order of tags in weak labels. Second, to utilize SLD in audio tagging, we propose to use CTC technique to train a CRNN and compare its performance with other common CRNN models in previous works. This paper is organized as follows, Section 1.2 introduces related works. Section 1.3 describes CRNN baseline. Section 1.4 describes CRNN-CTC with SLD. Section 1.5 describes dataset, experimental setup and results. Section 1.6 gives conclusions.

## 1.2 Related Work

Audio classification and detection have obtained increasing attention in recent years. There are many challenges for audio detection and tagging such as DCASE 2013 [3], DCASE 2016 [10] and DCASE 2017 [5].

In previous works in audio classification and tagging, Mel Frequency Cepstrum Coefficient (MFCC) and Guassian Mixture Model (GMM) is widely used in baseline system [3]. Recent methods include Deep Neural Networks (DNNs) [5], Convolution Neural Networks (CNNs) [11] and RNN [2], with inputs varying from Mel energy, spectrogram, MFCC to Constant Q Transform (CQT) [12].

Many methods described above rely on the bag of frames (BOF) model [13]. BOF is based on an assumption that tags occur in all frames, which is however not the case in practice. Some audio events like "gunshot" only happen a short time in audio clip. State-of-the-art audio tagging methods [14] transform waveform to the time-frequency (T-F) representation. T-F representation is treated as an image which is fed into CNNs. However, unlike image where the objects usually occupy a dominant part of the mage, in an audio clip audio events only occur a short time. To solve this problem, some attention models [15] for audio tagging and classification are applied to attend to the audio events and ignore the background sounds.

## 1.3 CRNN Baseline in Audio Tagging

CRNN has been successfully used in audio tagging [15]. First the waveforms of the audio recordings are transformed to time-frequency (T-F) representation such as log Mel spectrogram. Next Convolutional layers are applied on the T-F representation to extract high level features. Then, Bidirectional Gated Recurrent Units (BGRU) are adopted to capture the temporal context information. Finally, the output layer is a dense layer with the sigmoid activation function since it is a multi-class classification problem [2, 5, 10], the sigmoid activation function to predict probability of each sound events in the audio clip. Inspired by the good performance of CRNN in audio tagging [2, 15], we use CRNN as our baseline system in this paper.

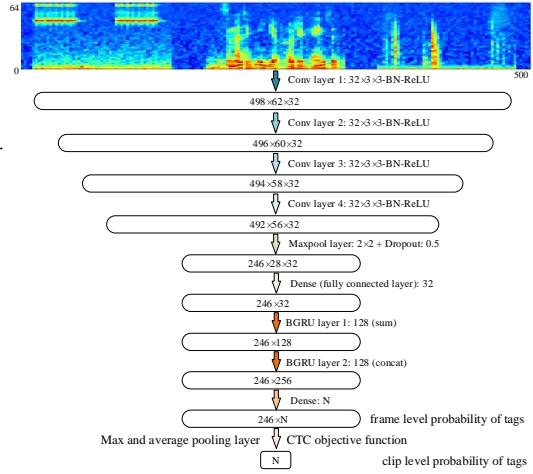

**Fig. 1.2** Model Structure.
BN: Batch Normalization.
ReLU: Rectified Linear Unit.
For baseline, CRMP and CRAP, N=16.
For CRNN-CCT, N=17 (16+1),
the extra '1' indicates the blank label.

An audio clip from real-life may contain more than one sound event, as environmental sound is often a mixture audio that come from multiple sound sources simultaneously. Thus the audio tagging task is a multi-label classification problem and a binary decision is made for each class [7]. In the training phase, the binary cross-entropy loss [16] is applied between the predicted probability of each tag and the ground truth tag in an audio clip. The loss can be defined as:

$$E = -\sum_{n=1}^{N} \left( P_n log Q_n + (1 - P_n) log (1 - Q_n) \right) \tag{1.1}$$

where $E$ is the binary cross-entropy, $Q_n$ and $P_n$ denote the predicted tags and reference tags sequence of the $n$-th audio clip, respectively. The batch size is represented by $N$.

In CRNN baseline, clip level probability of tags can be obtained from the last layer. However, there is no frame level information of each event in it. To obtain the probability of each event at each frame, a dense layer with the number of event classes, following the BGRU layer, as shown in Fig. 1.2. These frame level predictions can be used for sound event detection. To map the frame level tags to clip level tags, pooling layer was used. In training, the clip level predictions are compared against the weak labels of the audio clip to compute the loss function of model.

There are two pooling operations in Fig. 1.2, Max and Average Pooling. For CRNN with Max Pooling (CRMP) and CRNN with Average Pooling (CRAP), pooling performs down-sampling along time axis and transforms the frame level probability of tags to clip level tags, respectively. Max Pooling and Average Pooling as way of aggregation have been successfully used [17].

## 1.4 CRNN-CTC in Audio Tagging

As discussed before, strongly and weakly labelled data have their own drawbacks in audio tagging, so we propose sequential labelled data (SLD) and use CRNN-CTC to detect presence or absence of several sound events in SLD.

### 1.4.1 Sequential Labelled Data

Let $\mathcal{D}$ be a set of training examples drawn from audio dataset. Input space $\mathcal{X} = (\mathbb{R}^n)$ is the set of all sequences of $n$ dimensional vectors. Target space $\mathcal{Z} = L$ is the set of all sequences of labels over audio events. In general, we refer to elements of $L$ as label sequences or labellings [9]. Each example in $\mathcal{D}$ consists of a pair of sequences $(\boldsymbol{x}, \boldsymbol{z})$. The target sequence $\boldsymbol{z} = (z_1, z_2, ..., z_Q)$ is at most as long as input sequence $\boldsymbol{x} = (x_1, x_2, ..., x_T)$, i.e. $Q \leq T$. Since, the input and target sequences are not generally the same length, there is no a priori way of aligning them [9]. In the label sequence $\boldsymbol{z}$, the tags of the audio clip and sequence of tags are known, without knowing their occurrence time, that is, there is no starting/ending times of sound events. We refer to audio data labelled by label sequence as sequential labelled data (SLD).

In essence, SLD is a weakly labelled data with events sequence information. In audio tagging using SLD, we can use the model like CRNN described in section 1.3. However, there is no order information of sound events in predictions of baseline, CRMP and CRAP. And due to the drawbacks of Max Pooling and Average Pooling, predictions of CRMP in frame level often underestimates the occurrence probability of each events, while CRAP, in contrast, often overestimates them [18]. So we propose to use CRNN-CTC in audio tagging using SLD.

### 1.4.2 CRNN-CTC in Audio Tagging using SLD

CTC has achieved great success in speech recognition [8, 9]. In this section, we will show how to use CTC technique to train a CRNN in audio tagging using SLD.

CTC is a learning technique for sequence labelling, it shows a new way for training RNN with label unsegment sequences. In fact, CTC redefines the loss function of RNN [9] and allows RNN to be trained for sequence to sequence tasks, without requiring any prior alignment (i.e. starting or ending time of sound events) between the input and target sequences [8]. Thus, it is sufficient to train a CRNN using SLD with CTC technique. Given $y_t(k)$ is probability of observing label $k$ at time $t$ output by the last recurrent layer in CRNN, and $z_t$ is the ground-truth label, conventional loss function of RNN for a sequence $X$ of length $T$ is $L = -\sum_{t=1}^{T} log\, y_t(z_t)$, which is the negative logarithm of the joint probability of desired label sequence and its alignment. In audio tagging, we are only interested in label sequence, not the ground-truth alignment. Hence, we want to marginalize out the alignment.

CTC gives a solution to how to marginalize out the alignment. First, CTC adds an extra "blank" label (denoted by "-") to original label set $L$ [9]. Then, it defines a many-to-one mapping $\beta$ that transforms the alignment (i.e. the sequence of output labels at each time step, also called a path [9]) to label sequence. The mapping $\beta$ removes repeated labels from the path to a single one, then removes the "blank" labels. For example, $\beta(C - AT -) = \beta(-CC - -ATT) = CAT$, that is, path $'C - AT - '$ and $' - CC - -ATT'$ both map to the label sequence $'CAT'$.

The CTC objective function is defined as the negative logarithm of the total probability of all paths [8] that map to the ground-truth label sequence. The total probability can be found using dynamic programming algorithm [9] on the trellis shown in Fig. 1.3. On the $x$-axis is time steps, on the $y$-axis is "modified label sequence", that is target label sequence with blank labels added to the beginning and the end and inserted between every pair of labels. Given the length of modified label sequence is $L$ and $l_i$ denote $i$-th label. A effective path may start at either $l_1$ or $l_2$ and end at $l_{L-1}$ or $l_L$. At each time step, the path may i) stay at the same label; ii) move to the next label; iii) move to the label after the next if it is not a blank label different from the current label. Let $\alpha_t(s)$ be the total probability of $l_{1:s}$ at time $t$. Assuming conditional independence between $y_t(k)$ (i.e. probability of observing label $k$ at time $t$) across time steps, the $\alpha_t(s)$ can be calculated as follows:

$$\alpha_1(s) = \begin{cases} y_1(l_s) & s \leq 2 \\ 0 & s > 2 \end{cases} \tag{1.2}$$

$$\alpha_t(s) = [\alpha_{t-1}(s) + \alpha_{t-1}(s-1) + \delta_s \alpha_{t-1}(s-2)]y_t(l_s), t > 1 \tag{1.3}$$

where $\delta_s = 1$ if $l_s \neq l_{s-2}$, and terms that go past the start of the modified label sequence are zero. The sum of total probability of paths that map to original label sequence is $\alpha_T(L-1) + \alpha_T(L)$, and its negative logarithm is CTC loss function.

To decode the CTC output, there are several ways show in [9], and we use the simple best path decoding in this paper. This method is to select the label with the maximum probability at each frame, reduce adjacent repeating labels to a single one, and remove the blank labels. More details about CTC can be seen [9].

The output of CTC model is directly a label sequence corresponding the audio clip. The detailed structure of CRNN-CTC was shown before in Fig.1.2.

**Fig. 1.3** Trellis for computing CTC objective function [9] applied to the example labelling 'CAT'. Black circles represent labels, white circles represent blanks. Arrows signify allowed transitions.

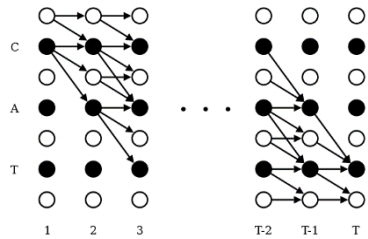

## 1.5 Experiments and Results

### 1.5.1 Dataset, Experiments Setup and Evaluation Metrics

We use the audio events in DCASE 2013 [3] to make SLD and evaluate the proposed method. There are 16 kinds of sound events in DCASE 2013 includes: *alert, clearthroat, cough, doorslam, drawer, keyboard, keys, knock, laughter, mouse, pageturn, pendrop, phone, printer, speech* and *switch*. We remixed these sound events to 10-second audio clips totaling 7.1 hours, where each audio clip contains no overlapped three or several sound events mixed with noise background.

For experimental setup, four-fold cross validation was used for model selection and parameter tuning. Dropout, batch-normalization and early stopping criteria are used in training phase to prevent over-fitting. The model is trained for maximum 1000 epochs with Adam optimizer with learning rate of 0.001.

To evaluate the results of audio tagging, we follow the metrics proposed in [17]. The results are evaluated by *precision, recall, F-score* [19] and Area Under Curve (*AUC*) [20]. To calculate these metrics, we need to count the number of: True Positive (*TP*), False Negative (*FN*) and False Positive (*FP*). Precision (*P*), Recall (*R*) and *F-score* [19] are defined as:

$$P = \frac{TP}{TP + FP}, \quad R = \frac{TP}{TP + FN}, \quad F = \frac{2P \cdot R}{P + R}. \quad (1.4)$$

To evaluate the True Positive Rate (*TPR*) versus False Positive Rate (*FPR*), the Receiver Operating Characteristic (*ROC*) curve was used [20]. *AUC* score is the area under the *ROC* curve which summarizes the *ROC* curve to a single number. Larger *P, R, F-score* and *AUC* indicates better performance.

### 1.5.2 Results

As the *AUC* score of audio tagging shown in Table 1.1, CRAP, CRMP and CRNN-CTC outperform baseline system. CRNN-CTC achieves an averaged *AUC* of 0.986.

**Table 1.1** *AUC* of Audio Tagging

| | alert | clear | cough | door | drawer | keybo | keys | knock | laugh | mouse | page | pendr | phone | print | speech | switch | avg. |
|---|---|---|---|---|---|---|---|---|---|---|---|---|---|---|---|---|---|
| Baseline | 0.609 | 0.627 | 0.674 | 0.691 | 0..690 | 0.569 | 0.702 | 0.816 | 0.617 | 0.668 | 0.693 | 0.662 | 0.654 | 0.862 | 0.550 | 0.625 | 0.669 |
| CRAP | 0.737 | 0.948 | 0.792 | 0.804 | 0.895 | 0.811 | 0.864 | 0.971 | 0.783 | 0.587 | 0.759 | 0.809 | 0.715 | 0.910 | 0.800 | 0.850 | 0.815 |
| CRMP | 0.959 | 0.970 | 0.915 | 0.875 | 0.953 | 0.735 | 0.918 | 0.973 | 0.883 | 0.835 | 0.892 | 0.936 | 0.892 | 0.985 | 0.887 | 0.922 | 0.908 |
| CRNN-CTC | **0.968** | **1.0** | **1.0** | **0.977** | **1.0** | **0.959** | **0.972** | **1.0** | **1.0** | **0.995** | **0.990** | **0.972** | **1.0** | **0.995** | **0.990** | **0.965** | **0.986** |

**Table 1.2** Averaged Stats of Audio Tagging

|  | *Precision* | *Recall* | *F-score* | *AUC* |
|---|---|---|---|---|
| Baseline | 0.687 | 0.371 | 0.482 | 0.669 |
| CRAP | 0.847 | 0.647 | 0.733 | 0.815 |
| CRMP | 0.933 | 0.827 | 0.877 | 0.908 |
| CRNN-CTC | **0.983** | **0.975** | **0.980** | **0.986** |

Table 1.2 shows the averaged statistic including *precision, recall, F-score* and *AUC* over 16 kinds of sound events, respectively, and CRNN-CTC performs better than other models. Fig. 1.4 shows the frame level predictions of models on example audio clip. In Fig. 1.4, CRNN-CTC predicts the tag sequence of audio clip, typically as a series of spikes [9]. Although the spikes align well with the actual position of sound events in audio clip, there is no time span information about these events.

In Fig. 1.4, CRMP produces wide peaks, indicating the onset and offset times of each event. That shows max pooling has ability to locate audio events, while average pooling seems to fail. The reason may be max pooling encourages the response for a single location to be high [18], for similar audio events which can obtain similar features. While average pooling in CRAP encourages all response to be high [18], the difference features of each event make it difficult to locate audio events.

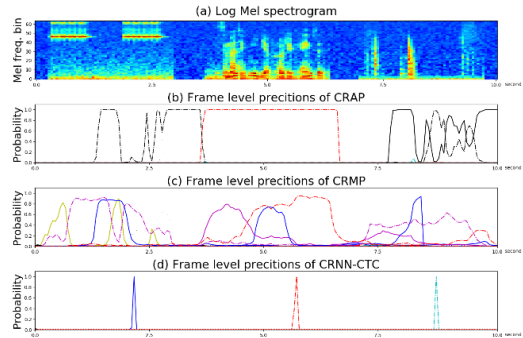

**Fig. 1.4** A Frame level predictions of CRAP (b), CRMP (c) and CRNN-CTC (d).
The ground-truth tag is "alert, speech, pageturn". Peaks are annotated with corresponding tag.

## 1.6 Conclusion

In this paper, we analyse the weakness of strongly and weakly labelled data, then propose SLD. To utilize SLD in audio tagging, we propose CRNN-CTC. In CRNN-CTC, CTC layer maps frame level tags to clip level tags, similar to the pooling layer. So we compare them. Experiments show CRNN-CTC outperforms CRAP, CRMP and baseline. The frame level predictions of models in Fig. 1.4 show CRNN-CTC predicts the presence/absence and tag sequence of events in the audio clip well.

# 1.7 References

1. G. Guo and Stan Z Li, "Content-based audio classification and retrieval by support vector machines," IEEE transactions on Neural Networks, vol. 14, no. 1, pp. 209–215, 2003
2. Y. Xu, Q. Kong and W. Wang, et al. "Large-scale weakly supervised audio classification using gated convolutional neural network," arXiv preprint arXiv: 1710.00343, 2017.
3. D. Stowell, D. Giannoulis and E. Benetos, et al. "Detection and classification of acoustic scenes and events," IEEE Transactions on Multimedia, vol. 17, no. 10, pp. 1733–1746, 2015
4. S. Dimitrov, J. Britz, B. Brandherm, and J. Frey, "Analyzing sounds of home environment for device recognition.," in AmI. Springer, 2014, pp. 1–16.
5. A. Mesaros, T. Heittola, A. Diment and B. Elizalde, et al. "DCASE 2017 challenge setup: Tasks, datasets and baseline system," in Proceedings of DCASE2017 Workshop.
6. Kong, Q., Xu, Y., Wang, W., Plumbley, M.D. A joint separation-classification model for sound event detection of weakly labelled data. arXiv preprint arXiv: 1711.03037, 2017.
7. A. Kumar and B. Raj, "Audio event detection using weakly labeled data," in Proceedings of the 2016 ACM on Multimedia Conference. ACM, 2016, pp. 1038–1047
8. A. Graves and N. Jaitly, "Towards end-to-end speech recognition with recurrent neural networks", in Proc. of ICML, 2014.
9. Graves A, Gomez F. Connectionist temporal classification: labelling unsegmented sequence data with recurrent neural networks[C]. International Conference on Machine Learning. ACM, 2006:369-376.
10. M Valenti, A Diment and G Parascandolo, et al., "DCASE 2016 acoustic scene classification using convolutional neural networks," Workshop on Detection and Classification of Acoustic Scenes and Events (DCASE 2016), Budapest, Hungary, 2016
11. Yoonchang Han and Kyogu Lee, "Acoustic scene classification using convolutional neural network and multiple-width frequency-delta data augmentation," arXiv preprint arXiv: 1607.02383, 2016.
12. Thomas Lidy and Alexander Schindler, "CQT-based convolutional neural networks for audio scene classification," in Workshop on Detection and Classification of Acoustic Scenes and Events (DCASE 2016), Budapest, Hungary, 2016
13. J. Ye, T. Kobayashi, M. Murakawa, and T. Higuchi, "Acoustic scene classification based on sound textures and events," in Proceedings of ACM on Multimedia Conference. ACM, 2015, pp. 1291–1294.
14. K. Choi, G. Fazekas, and M. Sandler, "Automatic tagging using deep convolutional neural networks," arXiv preprint arXiv: 1606.00298, 2016.
15. Y. Xu, Q. Kong, Q. Huang, W. Wang, and M. D. Plumbley, "Attention and localization based on a deep convolutional recurrent model for weakly supervised audio tagging," in INTERSPEECH. IEEE, 2017, pp. 3083–3087.
16. Farahnak-Ghazani, Fatemeh, and M. S. Baghshah. "Multi-label classification with feature-aware implicit encoding and generalized cross-entropy loss." Electrical Engineering IEEE, 2016:1574-1579.
17. Kong Q, Xu Y, Sobieraj I, et al. Sound Event Detection and Time-Frequency Segmentation from Weakly Labelled Data, arXiv preprint arXiv: 1804.04715, 2018.
18. Kolesnikov, Alexander, and C. H. Lampert. "Seed, Expand and Constrain: Three Principles for Weakly-Supervised Image Segmentation." European Conference on Computer Vision Springer International Publishing, 2016:695-711.
19. A. Mesaros, T. Heittola, and T. Virtanen, "Metrics for polyphonic sound event detection," Applied Sciences, vol. 6, no. 6, p. 162, 2016.
20. J. A. Hanley and B. J. McNeil, "The meaning and use of the area under a receiver operating characteristic (roc) curve." Radiology, vol. 143, no. 1, pp. 29–36, 1982.